\documentclass[11pt,a4paper]{article}
\pdfoutput=1
\usepackage{amsfonts}
\usepackage{jheppub}
\usepackage{amsmath}
\usepackage{amssymb}
\usepackage{xcolor}
\usepackage{float}
\usepackage{graphicx}%
\usepackage[normalem]{ulem}
\newcommand{\stkout}[1]{\ifmmode\text{\sout{\ensuremath{#1}}}\else\sout{#1}\fi}
\allowdisplaybreaks

\newcommand{\diff}{\mathrm{d}}

\title{Conformal Renormalisation of 8D Einstein Gravity}

\author[a]{Giorgos Anastasiou,}
\author[b]{Ignacio J. Araya,}
\author[c]{Nicolas Boulanger,}
\author[d]{Rodrigo Olea,}
\author[e]{Davide Rovere}

\affiliation[a]{Universidad Adolfo Iba\~nez, Facultad de Artes Liberales, Departamento de Ciencias,\\ Av. Diagonal Las Torres, 2640 Pe\~nalolen, Chile \vspace{0.1cm}}
\affiliation[b]{Universidad Andres Bello, Departamento de F\'isica y Astronom\'ia, \\  Facultad de Ciencias Exactas, Sazi\'e 2212, Piso 7, Santiago, Chile \vspace{0.1cm}}
\affiliation[c]{Service de Physique de l’Univers, Champs et Gravitation,
Universit\'e de Mons – UMONS,
20 place du Parc, 7000 Mons, Belgium \vspace{0.1cm}}
\affiliation[d]{Instituto de Física, Pontificia Universidad Católica de Valparaíso, Casilla 4059, Valparaíso, Chile.}
\affiliation[e]{Centre de Physique Théorique, CNRS, Institut Polytechnique de Paris, 91128 Palaiseau Cedex, France
 \vspace{0.1cm}}

\emailAdd{georgios.anastasiou@uai.cl}
\emailAdd{ignacio.araya@unab.cl}
\emailAdd{nicolas.boulanger@umons.ac.be}
\emailAdd{rodrigo{\_}olea{\_}a@yahoo.co.uk}
\emailAdd{davide.rovere@polytechnique.edu}

\date{March 2025}

\abstract{

We show that Holographic Renormalisation (HR) in eight 
dimensions is encoded 
in the unique conformal gravity theory that admits an Einstein 
sector with constant negative curvature.
We explicitly relate HR to Topological Regularisation (TR), 
where the latter prescribes 
to add the Euler term to the Einstein-Hilbert Lagrangian density, 
with a precise coefficient so as to ensure that the resulting 
density is polynomial in the anti de Sitter (AdS) curvature.
The polynomial is still asymptotically divergent.
We find that the aforementioned, unique conformal gravity action 
in 8D, reproduces the polynomial, together with extra boundary terms 
that cancel the divergent terms in the action.
We conjecture that, in arbitrary even dimension, 
HR is equivalent to conformally completing 
the Einstein-Hilbert action with negative cosmological constant, 
plus the Euler term with fixed coupling prescribed by TR.
}

\begin{document}

\maketitle

\section{Introduction}

\bigskip

In even-dimensional asymptotically anti-de Sitter gravity, the addition of a 
topological term of the Euler class \cite{Olea:2005gb,Miskovic:2008ck} 
may be rendered consistent by resorting to simple, compelling arguments 
based on black hole thermodynamics.

Indeed, even though topological terms can be locally written as boundary 
contributions, they are expected to modify the Noether charges of the theory 
\cite{Aros:1999id,Aros:1999kt}. Furthermore, if one thinks that black hole 
thermodynamics can be derived from a conservation law along the normal 
direction to the boundary, 
that implies that Smarr-type relations would be 
altered by the inclusion of the Euler term in the bulk  \cite{Olea:2005gb}.
Einstein-AdS gravity theory in $d+1$ dimensions is described by the action
\begin{equation}\label{EHaction}
I_{\mathrm{EH}}[g_{\mu\nu}]
=\frac{1}{16\pi G }\,\int\limits_{\mathcal{M}_{d+1}}d^{d+1}x\sqrt{-g} 
\,(R+d(d-1)/ \ell^2) \,,  
\end{equation}
where $\ell$ is the AdS$_{d+1}$ radius. The equations of motion, 
obtained upon varying this action with respect to the metric, 
read
\begin{equation}\label{eom}
   R_{\mu\nu} - \frac{1}{2}\,g_{\mu\nu}\,R -\frac{d\,(d-1)}{2\ell^2}\,g_{\mu\nu}= 0\,.
\end{equation}
We will call generic solutions to these field equations 
\emph{Einstein-AdS manifolds}.\footnote{Recall \cite{Eisenhart} that 
an Einstein manifold is a (pseudo)Riemannian 
manifold such that the traceless part of the Ricci tensor is zero, i.e., 
in dimension $n$ one has $R_{\mu\nu} - \frac{R}{n}\,g_{\mu\nu}=0\,$, 
where the scalar curvature $R$ does not need to be constant, in general.} 
Taking the trace and replacing the scalar curvature $R$ back in \eqref{eom}, 
one obtains the value of the 
(components of the) 
Ricci tensor of an 
Einstein-AdS manifold\footnote{With abuse of terminology, 
from now on we will identify a tensor with its components in a local chart.}:
\begin{equation}
   \left. R_{\mu\nu}\right|_{E}=-\frac{d}{\ell^2}\,g^{(E)}_{\mu\nu}\,,
\end{equation}
where here and in the following the symbol $|_E$ denotes the evaluation on Einstein-AdS manifolds, and $g^{(E)}_{\mu\nu}$ is a metric of an Einstein-AdS manifold, solution to \eqref{eom}. 

Let us recall the definition of the Schouten tensor:
\begin{equation}
    S_{\mu\nu} =\frac{1}{d-1}\,(R_{\mu\nu} - \frac{1}{2d}\,R\,g_{\mu\nu})\;.
\end{equation}
Note in passing that the traceless components of the Ricci and Schouten tensors 
are proportional to each other.
When evaluated on an Einstein-AdS manifold, the Schouten tensor reduces to
\begin{equation}
   \left. S_{\mu\nu}\right|_E = -\frac{1}{2\,\ell^2}\,g^{(E)}_{\mu\nu}\,,
\end{equation}
where the spacetime dimension dependence dropped out. 

One of the simplest solutions to \eqref{eom} is the Schwarzschild-AdS 
black hole which, in its Euclidean form, is written as
\begin{equation}
    ds^2=f^2(r)\,d\tau^2+f^{-2}(r)\,dr^2+r^2\,d\Omega^2_{d-1}\;,
\end{equation}
where $\tau=i t$ is the Euclidean time with period $\beta=1/T$ 
defined as the inverse of the Hawking temperature, 
the line element $d\Omega^2_{d-1}$ is the 
one of the sphere $\mathbb{S}^{d-1}$, 
and the function $f^2(r)$ is given by
\begin{equation}
    f^2(r)=1-\frac{2\,\omega_d\,G\, M}{r^{d-2}}\,+\frac{r^2}{\ell^2}\;,
\end{equation}
in terms of the  mass parameter $M$. 
The numerical factor $\omega_d$ is a matter of conventions, such that the 
black hole entropy is $S=\rm{Area}/4G$ in every dimension.

The Euclidean continuation of the bulk action $\tilde{I}_{\rm EH}\,$, evaluated 
on the static black hole with period $\beta$, gives rise to the expression
\begin{equation}
T\tilde{I}_{\rm EH}=\frac{d-2}{d-1}\,M-TS+\lim_{r\rightarrow \infty }\frac{\mathrm{Vol}(\mathbb{S}^{d-1})}{8\pi G }\frac{r^{d}}{\ell ^{2}}\,.
\end{equation}%
In the above formula there are two  fundamental problems:  an anomalous (Komar) factor for the internal energy in terms of the mass, and a divergent behaviour proper of asymptotically AdS spaces.

In every even bulk dimension ($d+1=2n$), the AdS gravity action may appear augmented by the Euler invariant
\begin{equation} \label{IEH+Euler}
    I[g_{\mu\nu}]=I_{\rm EH}[g_{\mu\nu}]
    +\alpha_{2n}\int\limits_{\mathcal{M}_{2n}}d^{2n}x 
\,\mathcal{E}_{2n}\,,
\end{equation}
which is defined as
\begin{equation}\label{EulerInv}
\mathcal{E}_{2n} = \frac{\sqrt{-g}}{2^n}\,\delta_{\nu_1\cdots\nu_{2n}}^{\mu_1\cdots\mu_{2n}}\,R^{\nu_1\nu_2}_{\mu_1\mu_2}\cdots R^{\nu_{2n-1}\nu_{2n}}_{\mu_{2n-1}\mu_{2n}}\;,
\end{equation}
where the symbol 
\begin{equation}
\delta_{\nu_1\cdots\nu_{2k}}^{\mu_1\cdots\mu_{2k}} = (2k)!\,
    \delta^{[\mu_1}_{\nu_1}\ldots \delta^{\mu_{2k}]}_{\nu_{2k}}
\end{equation}
using the strength-one antisymmetrisation convention. 
The topological character of the Euler term leaves the field equations unchanged, such that the coupling constant $\alpha_{2n}$ remains arbitrary at this stage. 
We will denote by $\tilde{\cal E}_{2n}$ the value of the Euler density evaluated on 
the Euclidean black hole solution. 
In the Euclidean sector of the theory, this topological contribution leads to 
a term proportional to the mass plus an infinite piece,

\begin{equation}
\alpha_{2n}\,T\int\limits_{\mathcal{M}_{2n}}d^{2n}x\,\tilde{\mathcal{E}}_{2n}=(-1)^n 16\pi G \frac{(2n-2)!}{\ell^{2n-2}}\alpha_{2n}\Bigg(\frac{M}{(2n-2)}-\lim_{r\rightarrow \infty }\frac{\mathrm{Vol}(\mathbb{S}^{2n-2})}{8\pi G }\frac{r^{2n-1}}{\ell ^{2}}\Bigg)\,,
\end{equation}
plus an additive constant proportional to the Euler characteristic of 
the horizon, which we did not write because it does not modify the 
First Law of Thermodynamics \cite{Olea:2005gb}.

As a consequence, the following standard thermodynamic relation 
for Schwarzschild-AdS black holes,  relating the Helmholtz free energy 
$F$ to the internal energy $U=M$, 
\begin{equation} \label{QSR}
F=T\tilde{I}=U-TS\,,
\end{equation}%
is recovered if the Euler coupling is fixed by \cite{Olea:2005gb}

\begin{equation} \label{alpha2n}
    \alpha_{2n}=  \frac{1}{16\pi G}\,\frac{(-1)^n}{n\,(2n-2)!}\,\ell^{2n-2}\;.
\end{equation}
The action \eqref{IEH+Euler} for the particular coupling in \eqref{alpha2n} 
has a number of remarkable features. Indeed, just for that value, 
it can be shown \cite{Miskovic:2009bm} that the action is identically zero 
for a spacetime that is locally AdS, i.e., for a metric which satisfies 
the relation
\begin{equation}
    R_{\mu \nu}{}^{\alpha \beta}+\frac{1}{\ell^2}\,
    (\delta^{\alpha}_{\mu} \delta^{\beta}_{\nu} - 
    \delta^{\alpha}_{\nu} \delta^{\beta}_{\mu})=0\;.
\end{equation}
This result suggests that the action is factorisable by the 
tensor appearing on the left-hand side of the above relation. 
Remarkably enough, when torsion vanishes, 
this tensorial quantity is the only nonvanishing part of the curvature for 
the AdS group $SO(d,2)$. 
On the other hand, the Weyl tensor provides a way to measure the 
deviation of the spacetime geometry away from a conformally-flat 
background, and in particular, from a maximally-symmetric space.
As a matter of fact, for Einstein-AdS spaces, 
these two notions of curvature coincide.
Let us define 
\begin{equation}
\label{Omega}
\Omega_{\mu \nu}{}^{\alpha \beta}:=R_{\mu \nu}{}^{\alpha \beta}
+\frac{1}{\ell^2}\,(\delta^{\alpha}_{\mu} \delta^{\beta}_{\nu} - 
    \delta^{\alpha}_{\nu} \delta^{\beta}_{\mu})\;.  
\end{equation}
This tensor coincides with the component of the $SO(2,d)$ 
curvature two-form along the Lorentz $SO(1,d)$ generators. 
The vanishing of torsion and of the Lorentz-valued two-form 
$\frac{1}{2}\,{\rm d}x^\mu\wedge {\rm d}x^\nu\,
\Omega_{\mu \nu}{}^{a b}\,M_{ab}$ 
implies that the geometry is the one of AdS$_{d+1}\,$.
When evaluated on Einstein-AdS manifolds,  
the tensor $\Omega_{\mu\nu}{}^{\alpha\beta}$ reduces to the 
Weyl tensor,  
\begin{equation}
\left.\Omega_{\mu\nu}{}^{\alpha \beta}\right|_E 
= W_{\mu \nu}{}^{\alpha \beta}\;,   
\label{omegaweyl}
\end{equation}
which is the $O(2n)$-irreducible component of the Riemann tensor that 
is left unconstrained by Einstein's equations \eqref{eom}.
For the appropriate choice \eqref{alpha2n} of coefficient $\alpha_{2n}$, 
it is remarkable that the action \eqref{IEH+Euler} can be written 
in the form \cite{Miskovic:2014zja}
\begin{equation} \label{polynomialWE}
    I[g_{\mu\nu}]=\,\frac{1}{16\pi G} \int\limits_{\mathcal{M}_{2n}}
    d^{2n}x\,\sqrt{-g}\,\sum_{k=2}^{n} a_{k} \Phi_{(k)}\;,
    \end{equation}
where $\Phi_{(k)}$ is the following symmetric polynomial of degree 
$k$ in $\Omega_{\mu\nu}{}^{\rho\sigma}$:
    \begin{equation} \label{Phi_E}
    \Phi_{(k)} := \delta_{\nu_1\cdots\nu_{2k}}^{\mu_1\cdots\mu_{2k}}\;
    \Omega_{\mu_1\mu_2}{}^{\nu_1\nu_2}\cdots \;
    \Omega_{\mu_{2k-1}\mu_{2k}}{}^{\nu_{2k-1}\nu_{2k}}\;.
    \end{equation} 
The coefficients $a_k$ are given by
\begin{equation}
a_k = \frac{(-1)^k}{2^{2k-1}}\frac{(2n-2k-1)!!}{k!(2n-3)!!}\;.
\end{equation}
%\begin{equation}\label{alpha_k}
%\textcolor{red}{a_k = \frac{(-1)^k\,(2\,n-1)\,n!\,(2\,n-2\,k)!}{2^{k-1}\,k!\,(2\,n)!\,(n-k)!}\,\ell^{2k-2}}
%\end{equation}
It is indeed remarkable that the above form of the AdS gravity action is 
polynomial in the tensor $\Omega$, 
and that the polynomial starts 
at second order. As a matter of fact, this is required by the following 
consistency condition: the thermodynamic relation \eqref{QSR} must hold 
for the ground state of the theory. While it is pretty obvious that $I$ 
is identically vanishing for the global AdS space, the use of Noether-Wald 
(NW) charges \cite{Wald:1993nt,Iyer:1994ys} is essential to check that 
one has ``$0=0$" identically. To be more precise, 
NW charges are defined in terms of the surface integral
\begin{align}\label{NWcharge}
    Q[\xi] = 2\int\limits_{\Sigma_\infty} E^{\mu\nu}_{\lambda\sigma}\nabla^\lambda\xi^\sigma \diff{\Sigma_{\mu\nu}}\,,
\end{align}
where $\xi=\xi^\mu\partial_\mu$ is a Killing vector, $E^{\mu\nu}_{\lambda\sigma}$ 
is the variational derivative of the Lagrangian with respect to the 
Riemann tensor, and $\Sigma_\infty$ is a codimension-2 asymptotic surface with 
area element $\diff{\Sigma_{\mu\nu}}$. Applied to the gravity action defined in 
\eqref{polynomialWE}, the NW method gives rise to conserved quantities which are 
at least linear in the Weyl tensor. 
Therefore, they are also zero for the gravitational vacuum configuration. 
On the other hand, as black hole entropy is the NW charge evaluated at the 
horizon, it is identically vanishing for the ground state of the theory, as well.

However, the addition of topological terms fails to provide a proper 
renormalisation scheme for generic solutions of Einstein-AdS gravity. Indeed, 
the gravity action for AAdS spaces contains divergent pieces when the boundary 
is not conformally flat, and we note that the geometry at the 
conformal boundary of the Schwarzschild-AdS black hole is conformally flat, 
so that the divergences vanish in that specific case.
%(which is the case of Schwarzschild-AdS black holes). 
Therefore, Topological Renormalisation seems to be at odds with thermodynamics 
of gravitational instantons in dimensions higher than four,  
and as a consequence, 
new counterterms are needed, which should be expressed in terms 
of the boundary Weyl tensor and derivatives thereof \cite{Anastasiou:2020mik}.

These new counterterms are switched on by conformal properties of the 
boundary and, a posteriori, may be seen as an ambiguity in the definition 
of the total action. In fact, while the bulk action is a polynomial in the 
$\Omega^{\mu\nu}{}_{\alpha\beta}$ tensor, additional boundary terms which 
depend on (derivatives of) $\Omega$ may be considered, 
which identically vanish on local AdS space.

In this work, further evidence on the deep connection between 
renormalisation of Einstein-AdS gravity and conformal invariants is provided, 
this time, by the 8D case. A unique combination of conformal invariants 
defines a Conformal Gravity theory that possesses an Einstein sector 
\cite{Boulanger:2025oli}. This allows to extend the framework of Conformal 
Renormalisation \cite{Anastasiou:2020mik,Anastasiou:2021tlv,Anastasiou:2023oro} 
to eight dimensions. New boundary terms, dictated by the Weyl invariance of the 
gravity action, are brought up in the Einstein sector and correctly renormalise 
it. As a matter of fact, these extra counterterms are written in terms of 
bulk-covariant quantities, such as the Weyl tensor, 
and the resulting action agrees with 
the one obtained by standard holographic techniques \cite{deHaro:2000vlm}, 
as we are going to show in the present paper.

The results are presented as follows: Section 2 shows how Topological 
Renormalisation breaks down in eight dimensions, as it is unable to render the 
Einstein-AdS action finite for AAdS spaces whose boundary is not conformally 
flat. The analysis is carried out resorting to the near-conformal boundary form 
of the metric \cite{Graham1985} and what 
it implies for the asymptotic form of the Weyl tensor. The divergent pieces in 
the holographic radial coordinate are identified in terms of the boundary 
Weyl tensor and its derivatives. 
Section 3 deals with the Chern form $B_7\,$, as that boundary 
contribution is locally equivalent to the Euler term in 8D. 
This part is intended to develop a reassuring argument, 
given by the explicit comparison between $B_7$ 
and standard counterterms given by Holographic Renormalisation. One then 
concludes that the mismatch is exactly the one found in Section 2. Section 4 
proposes a mechanism to achieve renormalisation in eight dimensions, by 
considering the Conformal Gravity theory that possesses an Einstein 
sector. 
That very precise conformal gravity action is given by the 
integral of the minimal combination of local conformal invariants 
in eight dimensions \cite{Boulanger:2004zfbis} 
such that the field equations are solved by Einstein-AdS spaces 
\cite{Boulanger:2025oli}. 
This result strongly suggests that renormalisation of 
Einstein-Hilbert gravity with negative cosmological constant 
in arbitrary even spacetime dimension 
is always encoded in a Weyl-invariant (or conformal) gravity theory.
The last section is devoted to concluding remarks, and three appendices 
give some technical details together with a short review of the 
Weyl-covariant derivative of \cite{Boulanger:2004eh}.

\section{Topological Terms and Einstein-AdS gravity in $8D$}

The Lagrangian appearing in \eqref{IEH+Euler}, using \eqref{polynomialWE} 
with $2n=8$, reads
\begin{equation}\label{TopRen8d}
R+\frac{42}{\ell^2} + \frac{\ell^6}{4\,6!\,\sqrt{-g}}\,\mathcal{E}_8 
= \frac{\ell^6}{80}\,\left(\frac{1}{\ell^4}\,\Phi_{(2)}-\frac{1}{36\,\ell^2}\,\Phi_{(3)}
+ \frac{1}{576}\,\Phi_{(4)}\right),
\end{equation}
where $\mathcal{E}_8$ is the eight-dimensional Euler 
invariant as in \eqref{EulerInv},
and the polynomials $\Phi_{(k)}$, $k=2,3,4$, are quadratic, cubic, 
and quartic polynomials in $\Omega$, defined in \eqref{Omega} 
according to \eqref{Phi_E}.
When evaluated on Einstein-AdS manifolds, they become the following polynomial in the Weyl tensor:
\begin{align}
\left.\Phi_{(2)}\right|_E &= 4\,W_{\alpha\beta\gamma\delta}\,W^{\alpha\beta\gamma\delta},\label{Phi2}\\
\left.\Phi_{(3)}\right|_E &= 32\,(2\,W_{\alpha\beta\gamma\delta}\,W^{\alpha\lambda\eta\beta}\,W_{\lambda}{}^{\gamma\delta}{}_{\eta} + W_{\alpha\beta\gamma\delta}\,W^{\gamma\delta\lambda\eta}\,W_{\lambda\eta}{}^{\alpha\beta}),\label{Phi3}\\
\left.\Phi_{(4)}\right|_E &= 96\,(-4\,W^{\alpha\beta\gamma\delta}\,W_{\alpha\beta}{}^{\epsilon\zeta}\,W_{\gamma\epsilon}{}^{\theta\iota}\,W_{\delta\zeta\theta\iota}
-16\,W^{\alpha\beta\gamma\delta}\,W_{\alpha}{}^{\epsilon}{}_{\gamma}{}^{\zeta}\,W_\beta{}^\theta{}_{\zeta}{}^\iota\,W_{\delta\iota\epsilon\theta} \,+\nonumber\\
& -64\, W^{\alpha\beta\gamma\delta}\,W_{\alpha\gamma}{}^{\epsilon\zeta}\,W_{\beta}{}^\theta{}_{\epsilon}{}^\iota\,W_{\delta\iota\zeta\theta} 
+ 8\,W^{\alpha\beta\gamma\delta}\,W_{\alpha}{}^{\epsilon}{}_{\gamma}{}^{\zeta}\,W_{\beta}{}^{\theta}{}_{\delta}{}^{\iota}\,W_{\epsilon\theta\zeta\iota} \,+\nonumber\\
& + 2\,W^{\alpha\beta\gamma\delta}\,W_{\alpha\beta}{}^{\epsilon\zeta}\,W_{\gamma\delta}{}^{\theta\iota}\,W_{\epsilon\theta\zeta\iota}
-16\,W^{\alpha\beta\gamma\delta}\,W_{\alpha\beta\gamma}{}^\epsilon \,W_{\delta}{}^{\zeta\theta\iota}\,W_{\epsilon\theta\zeta\iota} \,+\nonumber\\
& + W_{\alpha\beta\gamma\delta}\,W^{\alpha\beta\gamma\delta}\, W_{\epsilon\theta\zeta\iota}\,W^{\epsilon\zeta\theta\iota}).
\end{align}
Then, the topologically (partially) renormalised on-shell action reads
\begin{equation}\label{action+ct}
    \left.{I} [g_{\mu\nu}]\right|_{E} = \frac{\ell^6}{1280\pi G} \int\limits_{\mathcal{M}_8}d{^8x}\sqrt{-g}\left(\frac{1}{\ell^4} \left.\Phi_{\left(2\right)}\right|_E -\frac{1}{36 \ell^2} \left.\Phi_{\left(3\right)}\right|_E + \frac{1}{576} \left.\Phi_{\left(4\right)}\right|_E\right) \,.
\end{equation}
\iffalse
\begin{align}\label{action+ct}
    \left.{I} [g_{\mu\nu}]\right|_{E} &= \kappa\int\limits_{\mathcal{M}_8}d{^8x}\left.\left(\sqrt{-g}\,
    ({R}+\frac{42}{\ell^2}) + \frac{\ell^6}{4\,6!}\,{\mathcal{E}}_8\right)
    \right|_{E}\, \nonumber \\
    &=\frac{\kappa \ell^6}{80} \int\limits_{\mathcal{M}_8}d{^8x}\sqrt{-g}\left(\frac{1}{\ell^4} \left.\Phi_{\left(2\right)}\right|_E -\frac{1}{36 \ell^2} \left.\Phi_{\left(3\right)}\right|_E + \frac{1}{576} \left.\Phi_{\left(4\right)}\right|_E\right) \,.
\end{align}
\fi
Since $\left.\Phi_{\left(4\right)}\right|_E$ is the Pfaffian of the 
Weyl two-form, it generates no divergences. To study the asymptotic behaviour, 
we introduce the Fefferman-Graham (FG) expansion of the metric

\begin{equation}\label{gaussnormal}
    d{s^2} = \frac{\ell^2}{z^2}\left(d{z^2} + \bar{g}_{ij}(x,z)d{x^i}d{x^j}\right)\,,
\end{equation}
where $z$ denotes the holographic Poincaré coordinate and $\{x^i\}$ are the 
coordinates at the conformal infinity \cite{Graham1985}. 
The corresponding family of metrics at the boundary reads \cite{Graham:1999jg}
\begin{equation}
     \bar{g}_{ij}(x,z) = g_{(0)ij} + \frac{z^2 }{\ell^2}g_{(2)ij} + \frac{z^4}{\ell^4}g_{(4)ij} + \ldots \,.
     \label{FG-exp}
\end{equation}
where, for the boundary indices taking $d=2n-1$ values, we use Latin 
letters $i,j,k,\ldots$.
The Weyl squared term can be rewritten in the radial foliation as
\begin{equation}
W_{\mu \nu}{}^{\alpha \beta} W_{\alpha \beta}{}^{\mu \nu} 
= W_{mn}{}^{ij} W_{ij}{}^{mn} 
+ 4 W_{jz}{}^{iz} W_{iz}{}^{jz}+ 4 W_{mn}{}^{iz} W_{iz}{}^{mn} \,.
\end{equation}
Each one of the independent components of the Weyl tensor is expanded as 
\begin{align}\label{WeylAsymptotic}
W_{j z}{}^{i z} &= \frac{z^4}{3 \ell^2 } \mathcal{B}^{i}_{\left(0\right)j}+\mathcal{O} \left(z^6\right) \,, \notag \\
W_{j k}{}^{i z} &= - \frac{ z^3}{\ell^2}  \mathcal{C}_{(0)j k}^{i} + \mathcal{O} \left(z^5\right) \,, \notag \\
W_{k m}{}^{i j} &= \frac{z^2}{\ell^2} \mathcal{W}^{ij}_{\left(0\right) km}  -\frac{z^4}{\ell^2}\left[
\frac{2}{3}\,
\mathcal{B}_{(0)}^{[i}{}_{[k}
\delta^{j]}{}_{m]}
+
2\,
\mathcal{S}_{(0)}^{[i}{}_{n}\,
\mathcal{W}_{(0)}^{j]n}{}_{km}
-
\mathcal{S}_{(0)}^{n}{}_{[k}\,
\mathcal{W}_{(0)}^{ij}{}_{m]n}
+
D_{(0)}^{[i}
\mathcal{C}_{(0)}^{j]}{}_{km}
\right] \nonumber \\
& + \mathcal{O}\left(z^6\right) \,,
\end{align}
where ${\cal B}_{(0)}$, ${\cal C}_{(0)}$, ${\cal W}_{(0)}$, and ${\cal S}_{(0)}$, 
stand for the Bach, Cotton,
Weyl, and Schouten tensors of the boundary manifold, respectively, 
computed using the boundary metric $g_{(0)}$.
The boundary Levi-Civita covariant derivative associated with the metric
$g_{(0)}$ is denoted by the symbol $D_{(0)}$. 
Notice that the electric part of the Weyl tensor, given by the first component 
in Eq.~\eqref{WeylAsymptotic}, is of normalizable order only in four dimensions. 
In higher dimensions, one has to restrict the boundary metric 
to the equivalence class of conformally flat metrics 
in order to get the same fall-off. Conserved charges of this 
restricted class of spacetimes have been determined in 
Refs.~\cite{Ashtekar:1984zz,Ashtekar:1999jx,Jatkar:2014npa}.

Thus, for the quadratic term in the Weyl tensor, we get
\begin{align}
\left.{\Phi_{(2)}}\right|_E &= \frac{4z^4}{\ell^4} \mathcal{W}^{ij}_{\left(0\right)mn} \mathcal{W}^{mn}_{\left(0\right)ij} + \frac{8z^6}{\ell^4} \left[ \mathcal{S}^{m}_{\left(0\right)n} \mathcal{W}^{ js}_{\left(0\right)mi} \mathcal{W}^{ni}_{\left(0\right)js} + 2 \mathcal{C}^{ijm}_{\left(0\right)} \mathcal{C}_{\left(0\right)ijm} + \mathcal{W}_{\left(0\right) ijmn} D^{n}_{\left(0\right)} \mathcal{C}^{mij}_{\left(0\right)} \right] \nonumber\\
& +\mathcal{O} \left(z^8\right) \,, \label{Quadratic_Phi}
\end{align}
which can can be further simplified, noting that
\begin{equation}
\mathcal{W}_{(0)ijmn}\,D_{(0)}^{n}
\mathcal{C}_{(0)}^{mij}
=
D_{(0)}^{n}\!\left(
\mathcal{W}_{(0)ijmn}\,\mathcal{C}_{(0)}^{mij}
\right)
-
4\,\mathcal{C}_{(0)mij}\mathcal{C}_{(0)}^{mij}\,
\end{equation}
and neglecting total derivative terms at the boundary.
For the cubic term, we get
%\delta_{\mu_1\cdots\mu_6}^{\nu_1\cdots\mu_6}W^{\mu_1\mu_2}_{\nu_1\nu_2}W^{\mu_3\mu_4}_{\nu_3\nu_4}W^{\mu_5\mu_6}_{\nu_5\nu_6}
\begin{equation}
\left.{\Phi_{(3)}}\right|_E=\frac{z^6}{\ell^6}\delta_{i_1\cdots i_6}^{j_1\cdots j_6}\mathcal{W}^{i_1i_2}_{(0)j_1j_2}\mathcal{W}^{i_3i_4}_{(0)j_3j_4}\mathcal{W}^{i_5i_6}_{(0)j_5j_6}+\mathcal{O}\left(z^8\right) \,.
\end{equation}
Also, note that the FG expansion of the integration measure contributes to the 
divergences. 
Indeed, one contribution to the divergent parts of 
$\sqrt{-g}\,\left.\Phi_{(2)}\right|_E$
is
\begin{align}
\frac{\ell^8}{z^8}\sqrt{-g_{(0)}}
\left[
1-
\frac{z^2}{2}\mathcal{S}^{m}_{(0)m}+
\mathcal{O}(z^4)
\right]
\frac{4z^4}{\ell^4}\mathcal W_{(0)}^2
=
\frac{\ell^8}{z^8}\sqrt{-g_{(0)}}
\left[
\frac{4z^4}{\ell^4}\mathcal W_{(0)}^2
-
\frac{2z^6}{\ell^4}\mathcal{S}^{m}_{(0)m}\mathcal W_{(0)}^2
+
\mathcal{O}(z^8) 
\right],
\end{align}
obtained by expanding the density $\sqrt{-g}$ in the radial coordinate, 
and keeping only the first part of $\left.\Phi_{(2)}\right|_E$ 
from Eq.~\eqref{Quadratic_Phi}.
The term 
$-\frac{2z^6}{\ell^4}\mathcal{S}^{m}_{(0)m}\mathcal W_{(0)}^2$ 
coming from the integration measure combines with
$\frac{8z^6}{\ell^4} \mathcal{S}^{m}_{\left(0\right)n} \mathcal{W}^{ jl}_{\left(0\right)mi} \mathcal{W}^{ni}_{\left(0\right)jl}$ in Eq.~\eqref{Quadratic_Phi} to produce the contribution
$-\frac{z^6}{2}\left(\delta^{i_1 \ldots i_5}_{j_1 \ldots j_5} \mathcal{W}^{j_1 j_2}_{\left(0\right)i_1 i_2} \mathcal{W}^{j_3 j_4}_{\left(0\right)i_3 i_4} \mathcal{S}^{j_5}_{\left(0\right)i_5}\right)$.
In total, the divergent part of the action~\eqref{action+ct} at the conformal boundary ($z\rightarrow0$) reads 

\begin{align}\label{divergentpart}
 I_{\rm div}[g_{(0)ij}]&= \frac{\ell^6}{1280 \pi G} \int\limits_{\partial\mathcal{M}_8}d{^7x}\sqrt{-g_{\left(0\right)}} \left[\frac{4}{3 z^3} \mathcal{W}^{ij}_{\left(0\right)ms} \mathcal{W}^{ms}_{\left(0\right)ij} -\frac{1}{2z} \left(\delta^{i_1 \ldots i_5}_{j_1 \ldots j_5} \mathcal{W}^{j_1 j_2}_{\left(0\right)i_1 i_2} \mathcal{W}^{j_3 j_4}_{\left(0\right)i_3 i_4} \mathcal{S}^{j_5}_{\left(0\right)i_5} \right. \right. \nonumber \\
& \left. \left. + \frac{1}{18}\delta^{i_1 \ldots i_6}_{j_1 \ldots j_6} \mathcal{W}^{j_1 j_2}_{\left(0\right)i_1 i_2} \mathcal{W}^{j_3 j_4}_{\left(0\right)i_3 i_4} \mathcal{W}^{j_5 j_6}_{\left(0\right)i_5 i_6} 
+32\, \mathcal{C}^{ijm}_{\left(0\right)} \mathcal{C}_{\left(0\right)ijm}\right)\right] \,,
\end{align}
where total derivative terms at the boundary are neglected.

Notice that the divergences vanish identically only in the case of spacetimes with conformally flat boundary geometries, as explicitly shown in Ref.~\cite{Anastasiou:2020zwc}. 

This point justifies the fact that one can reproduce the thermodynamics of black holes with conformally flat boundaries by the addition of a single topological term in the bulk. However, this prescription is at odds if one considers AAdS solutions with nontrivial conformal properties at the boundary, as it is the case of gravitational instantons (e.g., Taub-NUT-AdS).

In the next sections, we perform standard holographic techniques to work out the form of the counterterms for arbitrary boundary geometries in eight-dimensional AdS gravity.
%This is necessary to show that the divergent pieces in Eq.~\eqref{divergentpart} are the same as the mismatch between Topological and Holographic Renormalisation.

\section{Comparison with Holographic Renormalisation}

At this point it is useful to perform an explicit comparison between Topological and Holographic Renormalisation. The two schemes agree for asymptotically conformally flat manifolds, but in the generic case a mismatch appears \cite{Anastasiou:2020zwc}. The corresponding boundary terms, in these two prescriptions, are written in a covariant  form for the codimension-1 metric of the radial foliation in Gauss-normal coordinates \cite{Henningson:1998gx,deHaro:2000vlm}
\begin{equation}\label{Gaussian}
ds^2=N^2(z)dz^2+h_{ij}(z,x)\,dx^idx^j \, .
\end{equation} 
The FG frame is readily recovered by setting

\begin{equation}\label{Induced_h}
N=\frac{\ell}{z}\,,\hspace{1cm} h_{ij}=\frac{\ell^2}{z^2}\,\bar{g}_{ij} \, .
\end{equation} 
In the case where the renormalisation is a consequence of standard holographic techniques, the total action is given by
\begin{equation}
I_{\text{ren}} =
\frac{1}{16 \pi G}
\int\limits_{\mathcal{M}_{d+1}}
d^{d+1}x\sqrt{-g} \left(R-2\Lambda\right) 
+\frac{1}{8 \pi G}\int\limits_{\partial \mathcal{M}_{d+1}}
d^{d}x\,\sqrt{-h}\,K 
+\int\limits_{\partial \mathcal{M}_{d+1}}d^{d}x\,\mathcal{L}_{ct}
\,,
\label{Sct}
\end{equation}
where $K$ is the trace of the extrinsic curvature
\begin{equation}
    K_{ij}=-\dfrac{1}{2N}\,\partial_{z}h_{ij}\,, \label{K}
\end{equation}
and the counterterms
\begin{align}\label{L-HRCounter}
8\pi G  \mathcal{L}_{ct} &= -\frac{d-1}{\ell }\sqrt{-h}-\frac{\ell \left(d-1\right)\sqrt{-h}}{%
\left(d-2\right)}\mathcal{S} + \frac{\ell^3 \sqrt{-h}}{2\left(d-4\right)} \delta^{ik}_{jl}\mathcal{S}^{j}_{i}\mathcal{S}^{l}_{k} \notag \\
&- \frac{\ell^{5}\sqrt{-h}}{\left(d-2\right)\left(d-4\right)\left(d-6\right)} \left(\mathcal{S}^i_j \mathcal{B}^j_i +\frac{d-4}{2} \delta^{ikm}_{jln}\mathcal{S}^{j}_{i}\mathcal{S}^{l}_{k} \mathcal{S}^{n}_{m} \right) +\mathcal{O}\left(\mathcal{R}^4\right)\,,
\end{align}
up to cubic terms in the curvature. These were originally computed in Refs.~\cite{deHaro:2000vlm,Anastasiou:2020zwc} but this time, we chose to write them down in terms of conformal tensors.

On the other hand, the Euler density of Eq.~\eqref{IEH+Euler} is equivalent to the $n$-th Chern form $B_{2n-1}$, by virtue of the Euler theorem for 
a $2n$-dimensional manifold $\mathcal{M}_{2n}$ with boundary
\begin{equation}
\int\limits_{\mathcal{M}_{2n}}d{^{2n}x}\,\mathcal{E}_{2n}=\left(4\pi\right)^{n}\,n!\,\chi\left(\mathcal{M}_{2n}\right) + \int\limits_{\partial \mathcal{M}_{2n}}d{^{2n-1}x}\,B_{2n-1} \,.
\label{Euler}
\end{equation}
They differ by a topological number, the Euler characteristic of the manifold. The corresponding boundary term is not purely intrinsic, but also extrinsic, and takes the form
\begin{align}
B_{2n-1} =-2n\sqrt{-h}\,\delta _{j_{1}\ldots j_{2n-1}}^{i_{1}\ldots i_{2n-1}}\,K_{i_{1}}^{j_{1}} \int \limits_{0}^{1}dt &\left (\frac{1}{2}\mathcal{R}_{i_{2}i_{3}}^{j_{2}j_{3}}\left (h\right ) -t^{2}K_{i_{2}}^{j_{2}}K_{i_{3}}^{j_{3}}\right ) \times  \nonumber 
\cdots \\ 
 & \times \left (\frac{1}{2}\mathcal{R}_{i_{2n-1}i_{2n-1}}^{j_{2n-1} j_{2n-1}}\left (h\right ) -t^{2}K_{i_{2n-1}}^{j_{2n-1}}K_{i_{2n-1}}^{j_{2n-1}}\right ) \,.
 \label{Chernform}
\end{align}
In order to compare the Topological Renormalisation with the form of the counterterms prescribed by Holographic Renormalisation, one has to add and subtract the Gibbons-Hawking-York term from Eq.~\eqref{IEH+Euler}. After expressing the Euler density in terms of the Chern form through Eq.~\eqref{Euler}, we get
\begin{equation}
\mathcal{L}_{K}\left (h ,K ,\mathcal{R}\right ) =\alpha_{2n}\,B_{2n-1} -\frac{1}{8 \pi G}\sqrt{ -h}\,K \,.
\label{kountertermsgen}
\end{equation}
Equivalently, the last expression can be written as
\begin{gather}
\mathcal{L}_{K} =\frac{\sqrt{ -h}}{8\pi G}\frac{\left ( -1\right )^{n-1}\ell ^{2n-2}}{\left (2n-2\right ) !}\delta _{j_{1}\ldots j_{2n-1}}^{i_{1}\ldots i_{2n-1}}K_{i_{1}}^{j_{1}}\Big[\int _{0}^{1}dt\left [\frac{1}{2}\mathcal{R}_{i_{2}i_{3}}^{j_{2}j_{3}} -t^{2}K_{i_{2}}^{j_{2}}K_{i_{3}}^{j_{3}}\right ] \times  \nonumber  \\
\cdots  \times  \left. \left[\frac{1}{2}\mathcal{R}_{i_{2n-1} i_{2n}}^{j_{2n-2} j_{2n}} -t^{2} K_{i_{2n-2}}^{j_{2n-2}}K_{i_{2n-1}}^{j_{2n-1}}\right ] +\frac{\left ( -1\right )^n}{\ell ^{2n-2}}\delta _{i_{2}}^{j_{2}} \cdots \delta _{i_{2n-1}}^{j_{2n-1}}\right]\,. \label{Lkounterodd}
 \end{gather}
Furthermore, the extrinsic curvature 
expands asymptotically as
\begin{equation}
K_{j}^{i}=\frac{1}{\ell }\,\delta _{j}^{i}+\ell \, \mathcal{S}_{j}^{i} -\frac{\ell^3}{\left(d-2\right) }\,\left[\frac{1}{d-4}\, \mathcal{B}^{i}_{j}+\mathcal{S}^{i}_{k}\mathcal{S}^{k}_{j}- \mathcal{S} \mathcal{S}^{i}_{j}-\frac{1}{2\left(d-1\right)}\left(\mathcal{S}^{a}_{k}\mathcal{S}^{k}_{a}-\mathcal{S}^2\right)\delta^{i}_{j}\right]+\mathcal{O}\left(\mathcal{R}^3\right)\,.
\label{K.expanded}
\end{equation}
All in all, Eq.~\eqref{Lkounterodd} expands up to cubic order in the boundary Riemann curvature as
\begin{equation}
\mathcal{L}_{K}= \mathcal{L}^{\left(0\right)}_{K }+\mathcal{L}^{\left(2\right)}_{K }+\mathcal{L}^{\left(4\right)}_{K}+\mathcal{L}^{\left(6\right)}_{K} +\mathcal{O}\left(\mathcal{R}^4\right) \,,
\end{equation}
where
\begin{align}
\mathcal{L}^{\left(0\right)}_{K }&=-\frac{\sqrt{ -h}}{8\pi G}\frac{d -1}{\ell } \,, \qquad 
\mathcal{L}^{\left(2\right)}_{K } =- \frac{\ell \sqrt{ -h}}{8\pi G}\frac{d -1}{d -2} \mathcal{S}\,, \\
\mathcal{L}^{\left(4\right)}_{K}&=\frac{\sqrt{ -h}}{8\pi G}\frac{\ell ^{3}}{2(d -4)}\left [\delta _{j_{1}j_{2}}^{i_{1}i_{2}} \mathcal{S}_{i_{1}}^{j_{1}}\mathcal{S}_{i_{2}}^{j_{2}} +\frac{1}{4\left(d -2\right)}\mathcal{W}_{ij}^{mn}\mathcal{W}_{mn}^{ij}\right ] \,, \\
\mathcal{L}^{\left(6\right)}_{K}&= -\frac{\sqrt{ -h}}{8\pi G}\frac{\ell ^{5}}{2(d -2)\left (d -6\right )}\left[\delta _{j_{1}j_{2}j_{3}}^{i_{1}i_{2}i_{3}}\mathcal{S}_{i_{1}}^{j_{1}}\mathcal{S}_{i_{2}}^{j_{2}}\mathcal{S}_{i_{3}}^{j_{3}} +\frac{1}{4}\delta _{j_{1}j_{2}j_{3}j_{4}}^{i_{1}i_{2}i_{3}i_{4}}\mathcal{W}_{i_{1}i_{2}}^{j_{1}j_{2}}\mathcal{S}_{i_{2}}^{j_{2}}\mathcal{S}_{i_{3}}^{j_{3}} \right.\nonumber  \\
& \left. +\frac{1}{16(d -4)}\delta _{j_{1}j_{2}j_{3}j_{4}j_{5}}^{i_{1}i_{2}i_{3}i_{4}i_{5}}\mathcal{W}_{i_{1}i_{2}}^{j_{1}j_{2}}\mathcal{W}_{i_{3}i_{4}}^{j_{3}j_{4}}\mathcal{S}_{i_{5}}^{j_{5}} +\frac{1}{192(d -4)}\delta _{j_{1}j_{2}j_{3}j_{4}j_{5}j_{6}}^{i_{1}i_{2}i_{3}i_{4}i_{5}i_{6}}\mathcal{W}_{i_{1}i_{2}}^{j_{1}j_{2}}\mathcal{W}_{i_{3}i_{4}}^{j_{3}j_{4}}\mathcal{W}_{i_{5}i_{6}}^{j_{5}j_{6}}\right] \,,
\end{align}
where $d=2n-1$. 
Notice that the two schemes agree in the first two terms in the series. However, additional terms mismatch appear in the quadratic order in the curvature and on. Indeed, we get that
\begin{align}
\mathcal{L}^{\left(4\right)}_{K}&=\mathcal{L}^{\left(4\right)}_{ct} + \frac{\sqrt{ -h}}{8\pi G}\frac{\ell ^{3}}{8 \left(d-2\right)\left(d -4\right)}\mathcal{W}_{ij}^{kl}\mathcal{W}_{kl}^{ij} \,,\nonumber \\
\mathcal{L}^{\left(6\right)}_{K}&=\mathcal{L}^{\left(6\right)}_{ct} + \frac{\sqrt{ -h}}{8\pi G}\frac{\ell ^{5}}{2 \left(d-2\right)\left(d -4\right) \left(d-6\right)}\left(2\mathcal{S}^i_j \mathcal{B}^j_i-\frac{d-4}{4} \delta _{j_{1}j_{2}j_{3}j_{4}}^{i_{1}i_{2}i_{3}i_{4}}\mathcal{W}_{i_{1}i_{2}}^{j_{1}j_{2}}\mathcal{S}_{i_{2}}^{j_{2}}\mathcal{S}_{i_{3}}^{j_{3}} \right. \nonumber \\
& \left. -\frac{1}{16}\delta _{j_{1}j_{2}j_{3}j_{4}j_{5}}^{i_{1}i_{2}i_{3}i_{4}i_{5}}\mathcal{W}_{i_{1}i_{2}}^{j_{1}j_{2}}\mathcal{W}_{i_{3}i_{4}}^{j_{3}j_{4}}\mathcal{S}_{i_{5}}^{j_{5}} -\frac{1}{192}\delta _{j_{1}j_{2}j_{3}j_{4}j_{5}j_{6}}^{i_{1}i_{2}i_{3}i_{4}i_{5}i_{6}}\mathcal{W}_{i_{1}i_{2}}^{j_{1}j_{2}}\mathcal{W}_{i_{3}i_{4}}^{j_{3}j_{4}}\mathcal{W}_{i_{5}i_{6}}^{j_{5}j_{6}}\right) \,.\label{counterdensities}
\end{align}
The asymptotic expansion of each of the terms given above reads
\begin{align}
\sqrt{ -h}\delta _{j_{1}j_{2}j_{3}j_{4}j_{5}j_{6}}^{i_{1}i_{2}i_{3}i_{4}i_{5}i_{6}}\mathcal{W}_{i_{1}i_{2}}^{j_{1}j_{2}}\mathcal{W}_{i_{3}i_{4}}^{j_{3}j_{4}}\mathcal{W}_{i_{5}i_{6}}^{j_{5}j_{6}} &=\frac{\ell \sqrt{ -g_{\left(0\right)}}}{z} \delta _{j_{1}j_{2}j_{3}j_{4}j_{5}j_{6}}^{i_{1}i_{2}i_{3}i_{4}i_{5}i_{6}}\mathcal{W}_{\left(0\right)i_{1}i_{2}}^{j_{1}j_{2}}\mathcal{W}_{\left(0\right)i_{3}i_{4}}^{j_{3}j_{4}}\mathcal{W}_{\left(0\right)i_{5}i_{6}}^{j_{5}j_{6}}+\mathcal{O}(z) \,, \nonumber\\
\sqrt{ -h} \delta _{j_{1}j_{2}j_{3}j_{4}j_{5}}^{i_{1}i_{2}i_{3}i_{4}i_{5}}\mathcal{W}_{i_{1}i_{2}}^{j_{1}j_{2}}\mathcal{W}_{i_{3}i_{4}}^{j_{3}j_{4}}\mathcal{S}_{i_{5}}^{j_{5}} &= \frac{\ell \sqrt{ -g_{\left(0\right)}}}{z} \delta _{j_{1}j_{2}j_{3}j_{4}j_{5}}^{i_{1}i_{2}i_{3}i_{4}i_{5}}\mathcal{W}_{\left(0\right)i_{1}i_{2}}^{j_{1}j_{2}}\mathcal{W}_{\left(0\right)i_{3}i_{4}}^{j_{3}j_{4}}\mathcal{S}_{\left(0\right)i_{5}}^{j_{5}}+\mathcal{O}(z) \,, \nonumber \\
\sqrt{ -h} \delta _{j_{1}j_{2}j_{3}j_{4}}^{i_{1}i_{2}i_{3}i_{4}}\mathcal{W}_{i_{1}i_{2}}^{j_{1}j_{2}}\mathcal{S}_{i_{2}}^{j_{2}}\mathcal{S}_{i_{3}}^{j_{3}} &= \frac{\ell \sqrt{ -g_{\left(0\right)}}}{z} \delta _{j_{1}j_{2}j_{3}j_{4}}^{i_{1}i_{2}i_{3}i_{4}}\mathcal{W}_{\left(0\right)i_{1}i_{2}}^{j_{1}j_{2}}\mathcal{S}_{\left(0\right)i_{2}}^{j_{2}}\mathcal{S}_{\left(0\right)i_{3}}^{j_{3}}+\mathcal{O}(z) \,, \nonumber \\
\sqrt{ -h} \mathcal{S}_{j}^{i}\mathcal{B}_{i}^{j} &= \frac{\ell \sqrt{ -g_{\left(0\right)}}}{z} \mathcal{S}_{\left(0\right)j}^{i}\mathcal{B}_{\left(0\right)i}^{j}+\mathcal{O}(z) \,, \nonumber \\
\sqrt{ -h} \mathcal{W}_{ij}^{kl}\mathcal{W}_{kl}^{ij} &= \frac{\ell^3 \sqrt{ -g_{\left(0\right)}}}{z^3} \left[\mathcal{W}_{\left(0\right)ij}^{kl}\mathcal{W}_{\left(0\right)kl}^{ij} +z^{2} \left(4 \mathcal{S}_{\left(0\right)j}^{i} \mathcal{S}_{\left(0\right)b}^{a} \mathcal{W}_{\left(0\right)ia}^{jb} \right. \right. \nonumber \\
&\left. \left. - \frac{1}{8}\delta _{j_{1}j_{2}j_{3}j_{4}j_{5}}^{i_{1}i_{2}i_{3}i_{4}i_{5}}\mathcal{W}_{\left(0\right)i_{1}i_{2}}^{j_{1}j_{2}}\mathcal{W}_{\left(0\right)i_{3}i_{4}}^{j_{3}j_{4}}\mathcal{S}_{\left(0\right)i_{5}}^{j_{5}} +2 \mathcal{W}_{\left(0\right) klim} D^{m}_{\left(0\right)} \mathcal{C}^{ikl}_{\left(0\right)}\right)\right] +\mathcal{O}(z)\,. \nonumber
\end{align}
Considering the expressions for the mismatch terms given in Eq.~\eqref{counterdensities}, together with the asymptotic expansions presented above, one can see that
\begin{equation}
\int\limits_{\partial \mathcal{M}_8} d^7x\,
\left(
\mathcal{L}_{ct}-\mathcal{L}_{K}
\right)
=
- I_{\rm div} \left[g_{\left(0\right)ij}\right]+\mathcal{O}(z)\,,
\end{equation}
where total derivatives at the conformal boundary have been dropped.

Therefore, we have proven that
the set of divergent contributions which comes from the addition of the Euler term to the Einstein action $I_{\rm div}$ in Eq.~\eqref{divergentpart} is the same as the mismatch with Holographic Renormalisation.

One could try to patch up the gravitational action with additional 
boundary terms. In particular, one may be interested in bulk-covariant 
expressions. As mentioned above, there is an ambiguity in the total action 
(Einstein plus Euler term) which is given by surfaces terms which depend 
on the bulk Weyl tensor.
These criteria restrict 
the total derivatives to be of the form
\begin{equation}\label{BndTerms}
\Box^2\Phi_{\left(2\right)}\,, \quad
\Box \Phi_{\left(3\right)}\,, \quad
\frac{1}{\ell^2}\,\Box \Phi_{\left(2\right)}\,,
\end{equation}
whose 
divergent behaviour in the asymptotic FG expansion is given by
\begin{align}
\Box^2\left.\Phi_{(2)}\right|_E
&=
\frac{576z^4}{\ell^8}\mathcal I_2
+
\frac{z^6}{\ell^8}
\left[
288\mathcal Q_2
+
288\mathcal{S}^{m}_{(0)m}\mathcal I_2
-
72D_{(0)}^m D_{(0)m}\mathcal I_2
\right]
+
\mathcal{O}(z^8)\,,\\
\Box\left.\Phi_{(3)}\right|_E
&=
-\frac{6z^6}{\ell^8}\mathcal I_3
+
\mathcal{O}(z^8)\,,\\
\Box\left.\Phi_{(2)}\right|_E
&=
-\frac{48z^4}{\ell^6}\mathcal I_2
+
\frac{z^6}{\ell^6}
\left[
4 D_{(0)}^m D_{(0)m}\mathcal I_2
-
16\mathcal{S}^{m}_{(0)m}\mathcal I_2
-
48\mathcal Q_2
\right]
+
\mathcal{O}(z^8)\,,
\end{align}
where we defined for convenience the auxiliary variables
\begin{align}
\mathcal I_2 &:= 
\mathcal W^{ij}_{(0)mn}\mathcal W^{mn}_{(0)ij}\,,\\
\mathcal I_3 &:=
\delta^{i_1\cdots i_6}_{j_1\cdots j_6}
\mathcal W^{j_1j_2}_{(0)i_1i_2}
\mathcal W^{j_3j_4}_{(0)i_3i_4}
\mathcal W^{j_5j_6}_{(0)i_5i_6}\,,\\
\mathcal Q_2 &:= 
\mathcal S^{m}_{(0)n}
\mathcal W^{jl}_{(0)mi}
\mathcal W^{ni}_{(0)jl}
-
2\mathcal C^{ijm}_{(0)}\mathcal C_{(0)ijm}
+
D^{n}_{(0)}\left(\mathcal W_{(0)ijmn}
\mathcal C^{mij}_{(0)}\right)\;.
\end{align}
Interestingly enough, there is a unique combination of these three terms that cancels the divergences of Eq.~\eqref{divergentpart}, up to total derivatives at the conformal boundary of the form $D_{(0)}^m D_{(0)m}\mathcal{I}_2$ and $ D^{n}_{(0)}\left(\mathcal W_{(0)ijmn}
\mathcal C^{mij}_{(0)}\right)$. Therefore, this particular prescription for renormalised action, reads
\begin{align}
   \left.I_{\rm ren} [g_{\mu\nu}]\right|_E
    =\frac{\ell^6}{1280 \pi G} \int\limits_{\mathcal{M}_8}d{^8x}\,\sqrt{-g}\,&\bigg{[}\frac{1}{\ell^4} \,\left.\Phi_{\left(2\right)}\right|_E -\frac{1}{36 \ell^2}\,\left.\Phi_{\left(3\right)}\right|_E + \frac{1}{576} \,\left.\Phi_{\left(4\right)}\right|_E\nonumber\\
    +\,&\frac{1}{72}\,\left.\Box^{2} \Phi_{\left(2\right)}\right|_E -\frac{1}{216}\,\Box \left.\Phi_{\left(3\right)}\right|_E +\frac{1}{ 4 \ell ^{2}}\, \Box \left.\Phi_{\left(2\right)}\right|_E\bigg{]} \,.\label{BulkBndAction}
\end{align}
In the next section, we show that this form of renormalised action appears in the Einstein sector of a specific 8D Conformal Gravity theory.

\section{$8D$ Conformal Gravity with an Einstein Sector}

The most general conformal gravity action 
in $8D$ that admits an Einstein sector was found in Ref.~\cite{Boulanger:2025oli},
\begin{equation}\label{8dConfActionEinstein}
I_{\rm BR}[g_{\mu\nu}] = 
\int\limits_{\mathcal{M}_8}d^8 x\,\sqrt{-g}\,{L}_8\;,
\end{equation}
where the Lagrangian density $\mathcal{L}_8 = \sqrt{-g}\,{L}_8$ 
of this gravity theory is a given combination of Weyl-invariant 
8D scalar densities \cite{Boulanger:2004zfbis}:
\begin{equation}\label{J8dConfActionEinstein}
\mathcal{L}_8 = -\tfrac{1}{4}\,(\mathcal{I}_1 - \tfrac{23}{25}\,
\mathcal{I}_2 -\tfrac{21}{25}\,I_3 - 39\,\mathcal{I}_6 
+ \tfrac{2}{5}\,\mathcal{I}_7 - \tfrac{7}{2}\,\mathcal{I}_8 
+ \tfrac{124}{5}\,\mathcal{I}_9 + \tfrac{9}{20}\,\mathcal{I}_{10} 
-4\,\mathcal{I}_{11} - \tfrac{568}{5}\,\mathcal{I}_{12})\;.
\end{equation}
Writing $\mathcal{I}_i = \sqrt{-g}\,I_i$,  the scalars 
$I_i$ where $i\in \{6, \ldots ,12\}\,$ provide a 
basis\footnote{We profit from this discussion to point out that the basis 
used in \cite{Boulanger:2025oli} is actually the one given here 
in \eqref{I6}--\eqref{I12}, and \emph{not} the one provided in 
the Eqs. (5.1)--(5.7) of \cite{Boulanger:2025oli}.
We also note a typo in the paper \cite{Boulanger:2004zfbis} 
where the invariant ${\cal I}_{10}$ therein should change sign to match 
with the basis given in \cite{Fulling:1992vm}.
We thank Zhenya Skvortsov for discussions about this point.} 
of the seven possible parity-even scalars that are quartic in the 
undifferentiated Weyl tensor \cite{Fulling:1992vm}
\begin{align}
I_6 &= W^{\alpha\beta\gamma\delta}
\,W_{\alpha\beta\gamma\delta}
\,W^{\mu\nu\rho\sigma}
\,W_{\mu\nu\rho\sigma}\;,\label{I6}\\
I_7 &= W^{\alpha\beta\gamma\delta}
\,W_{\alpha\beta\gamma}{}^{\mu}
\,W^{\nu\rho\sigma}{}_{\delta}
\,W_{\nu\rho\sigma\mu}\;,\label{I7}\\
I_8 &= W^{\alpha\beta\gamma\delta}
\,W_{\alpha\beta}{}^{\mu\nu}
\,W_{\mu\nu}{}^{\rho\sigma}
\,W_{\gamma\delta\rho\sigma}\;,\label{I8}\\
I_9 &= W^{\alpha\beta\gamma\delta}
\,W_{\alpha\beta}{}^{\mu\nu}
\,W_{\gamma\mu}{}^{\rho\sigma}
\,W_{\delta\nu\rho\sigma}\;,\label{I9}\\
I_{10} &= W^{\alpha\beta\gamma\delta}
\,W_{\alpha\beta}{}^{\mu\nu}
\,W_{\gamma}{}^{\rho}{}_{\mu}{}^{\sigma}
\,W_{\delta\rho\nu\sigma}\;,\label{I10}\\
I_{11} &= W^{\alpha\beta\gamma\delta}
\,W_{\alpha}{}^{\mu}{}_{\gamma}{}^{\nu}
\,W_{\mu}{}^{\rho}{}_{\nu}{}^{\sigma}
\,W_{\beta\rho\delta\sigma}\;,\label{I11}\\
I_{12} &= W^{\alpha\beta\gamma\delta}
\,W_{\alpha}{}^{\mu}{}_{\gamma}{}^{\nu}
\,W_{\mu}{}^{\rho}{}_{\beta}{}^{\sigma}
\,W_{\nu\rho\delta\sigma}\;.\label{I12}
\end{align}
There are also five independent non-trivial Weyl-invariant scalar densities 
that involve derivatives of the Weyl tensor \cite{Boulanger:2004zfbis}. 
The first three of these, $\sqrt{-g} \,I_j\,$, $j\in \{1, 2, 3\}\,$, 
are given by
\begin{align}
I_1 =\;& W^{\rho\gamma\mu\sigma}\,
{\cal D}^\alpha{\cal D}_\alpha{\cal D}^\beta
{\cal D}_\beta W_{\rho\gamma\mu\sigma}\,
+\tfrac{48}{25}\,{\cal D}_{\beta}W^\beta{}_{\gamma\mu\alpha}
{\cal D}^\alpha{\cal D}_\alpha{\cal D}_\rho W^{\rho\gamma\mu\alpha} \,
\nonumber \\
& +2\,{\cal D}_\alpha W_{\mu\beta\gamma\nu}\,
{\cal D}_\rho {\cal D}^\rho {\cal D}^\alpha W^{\mu\beta\gamma\nu} 
+ \tfrac{42}{125}\,{\cal D}^{\alpha}{\cal D}^{\beta}W_{\gamma\alpha\beta\mu}\,
{\cal D}_\nu {\cal D}_{\rho}W^{\gamma\nu\rho\mu} \,
\nonumber \\
& +\tfrac{9}{10}\,{\cal D}^\gamma {\cal D}_\gamma W_{\alpha\mu\nu\beta}\,
{\cal D}^\rho {\cal D}_\rho W^{\alpha\mu\nu\beta}
+\tfrac{3}{5}\,{\cal D}_\alpha  {\cal D}_\beta W_{\nu\gamma\mu\rho}\,
{\cal D}^\alpha {\cal D}^\beta W^{\nu\gamma\mu\rho} \,
\nonumber \\
& +\tfrac{96}{125}\,{\cal D}_\alpha {\cal D}_\gamma W^\gamma{}_{\mu\nu\beta}\,
{\cal D}^\alpha {\cal D}_\rho W^{\rho\mu\nu\beta} 
+\tfrac{74}{25}\,
W_\beta{}^{\alpha\gamma\mu}\,W_{\nu\alpha\gamma\mu}\,
{\cal D}^{\rho} {\cal D}^\sigma W^\beta{}_{\rho\sigma}{}^{\nu} \,
\nonumber \\
& + \tfrac{208}{5}\,W_{\mu\beta\gamma\alpha}\,W_\sigma{}^{\nu\rho\alpha}\,
{\cal D}^\beta {\cal D}^\gamma W^\mu{}_{\nu\rho}{}^{\sigma} 
-8\,W_\alpha{}^\gamma{}_\beta{}^\mu\,W^\alpha{}_\nu{}^\beta{}_\rho\,
{\cal D}^\sigma {\cal D}_\sigma W^\nu{}_\gamma{}^\rho{}_{\mu} \,
\nonumber \\
& +\tfrac{16}{5}\,W_{\alpha\gamma\mu\rho}\,
W_{\beta\nu}{}^{\alpha\gamma}\,{\cal D}_\sigma 
{\cal D}^\sigma W^{\beta\nu\mu\rho}
-\tfrac{144}{25}\,W^\gamma{}_\alpha{}^\mu{}_\beta\,
{\cal D}^\rho W_{\rho\gamma}{}^\nu{}_\mu{}\,
{\cal D}^\sigma W_\sigma{}^\alpha{}_\nu{}^{\beta} \,
\nonumber \\
& +\tfrac{104}{5}\,
W_\alpha{}^\gamma{}_\beta{}^\mu\,{\cal D}^\alpha W^\beta{}_\mu{}^{\sigma\nu}\,
{\cal D}^\rho W_{\rho\gamma\sigma\nu}
-\tfrac{88}{25}\,W_{\alpha\beta\gamma\mu}\,
{\cal D}^\rho W_{\rho\nu}{}^{\alpha\beta}\,
{\cal D}^\sigma W_\sigma{}^{\nu\gamma\mu}\;,
 \label{I1}\\
I_2 =\;& W_\beta{}^{\alpha\gamma\mu}\,W_{\nu\alpha\gamma\mu}\,
{\cal D}^\rho {\cal D}^\sigma W^\beta{}_{\rho\sigma}{}^{\nu}
+5\,W_{\alpha\gamma\mu\rho}\,W_{\beta\nu}{}^{\alpha\gamma}\,
{\cal D}^\sigma {\cal D}_\sigma W^{\beta\nu\mu\rho} \,
 \nonumber \\
& +5\,W_{\alpha\beta\gamma\mu}\,{\cal D}_\nu W^{\alpha\beta\rho\sigma}\,
{\cal D}^\nu W^{\gamma\mu}{}_{\rho\sigma}
+\tfrac{12}{5}\,W_{\alpha\beta\gamma\mu}\,
{\cal D}^\rho W_{\rho\nu}{}^{\alpha\beta}\,
{\cal D}^\sigma W_\sigma{}^{\nu\gamma\mu}\;,
\label{I2}\\
I_3 =\;& W_\beta{}^{\alpha\gamma\mu}\,W_{\nu\alpha\gamma\mu}\,
{\cal D}^\rho {\cal D}^\sigma W^\beta{}_{\rho\sigma}{}^{\nu}
-20\,W_\alpha{}^\gamma{}_\beta{}^\mu\,W^\alpha{}_\nu{}^\beta{}_\rho\,
{\cal D}^\sigma {\cal D}_\sigma W^\nu{}_\gamma{}^\rho{}_{\mu} \,
\nonumber \\
& -\tfrac{48}{5}\,W^\gamma{}_\alpha{}^\mu{}_\beta\,
{\cal D}^\rho W_{\rho\gamma}{}^\nu{}_\mu \, 
{\cal D}^\sigma W_\sigma{}^\alpha{}_\nu{}^{\beta}
-20\,W^\alpha{}_\mu{}^\gamma{}_\beta\, 
{\cal D}_\nu W^\mu{}_\rho{}^\beta{}_{\sigma}\,
{\cal D}^\nu W^\rho{}_\alpha{}^\sigma{}_\gamma\;.\label{I3}
\end{align}
where the Weyl-covariant derivative ${\cal D}$ 
introduced in Ref.~\cite{Boulanger:2004eh} is explained in 
Appendix \ref{sec:Weyl-covariant}.
As was found in Refs.~\cite{Chen:2024kuw} and \cite{Case:2024oih}, 
the other two Weyl-invariant scalar densities are total divergences. 
They can be written in terms of the Weyl tensor 
as \cite{Boulanger:2025oli}  $\sqrt{-g}\,\,\nabla_\alpha\,J_{(i)}^\alpha(W)\,$, 
$i=1,2\,$, where
\begin{align}
J_{(1)}^\alpha =&\, -\tfrac{1}{5}\,W_{\beta\gamma\delta}{}^{\sigma}\,W^{\beta\gamma\delta\epsilon}\,
\nabla^\rho W^{\alpha}{}_{\epsilon\sigma\rho}
+W^{\alpha\beta\gamma\delta}\,W_{\beta}{}^{\epsilon\sigma\rho}\,
\nabla_\rho W_{\gamma\delta\epsilon\sigma}
\label{J1} \nonumber \\
& -\tfrac{4}{15}\,
W^{\alpha\beta\gamma\delta}\,W_\beta{}^\epsilon{}_\gamma{}^\sigma\,
\nabla^{\rho}W_{\delta\epsilon\sigma\rho}\,
+\tfrac{8}{15}\,W^{\alpha\beta\gamma\delta}\,W_\beta{}^\epsilon{}_\gamma{}^\sigma\,
\nabla^{\rho}W_{\epsilon\sigma\delta\rho}
\nonumber \\
& +\tfrac{4}{15}\,W^{\alpha\beta\gamma\delta}\,W_\beta{}^\epsilon{}_\gamma{}^\sigma\,
\nabla^{\rho}W_{\delta\sigma\epsilon\rho}\;,
\\
J_{(2)}^\alpha =&\; 
W^{\alpha\beta\gamma\delta}\,W_\gamma{}^{\epsilon\sigma\rho}\,
\nabla_{\rho}W_{\beta\epsilon\delta\sigma}-W^{\alpha\beta\gamma\delta}\,
W_{\beta}{}^{\epsilon\sigma\rho}\, \nabla_{\rho} W_{\gamma\delta\epsilon\sigma}\,
\nonumber\\
& -\tfrac{2}{5}\,W^{\alpha\beta\gamma\delta}\,W_\beta{}^\epsilon{}_\gamma{}^\sigma\,
\nabla^{\rho}W_{\epsilon\sigma\delta\rho}
-\tfrac{2}{5}\,
W^{\alpha\beta\gamma\delta}\,W_\beta{}^\epsilon{}_\gamma{}^\sigma\,
\nabla^{\rho}W_{\delta\sigma\epsilon\rho}\;.
\label{J2}
\end{align}
Since they are total divergences, they do not change the equations of motion. 
The freedom one has in adding an arbitrary combination of these two invariants 
to the action \eqref{8dConfActionEinstein} is a novelty of the eight-dimensional 
case, absent in the lower-dimensional cases.

For the sake of completeness, we mention that two of the five 
non-trivial invariants of Refs.~\cite{Boulanger:2004zfbis},
namely ${\cal I}_i = \sqrt{-g}\,I_i$, $i=4,5$, 
can be expressed in terms of the remaining three non-trivial ones plus 
the seven undifferentiated ones of Eqs.~\eqref{I6}--\eqref{I12} 
and \eqref{I1}--\eqref{I3} respectively, and in terms of 
the total derivatives obtained from Eqs.~\eqref{J1} and \eqref{J2}, i.e.
\begin{align}
I_4 &= \tfrac{1}{40}\,I_2 - \tfrac{1}{40}\,I_3 + \tfrac{25}{3}\,I_6 
+ \tfrac{8}{3}\,I_7 + \tfrac{2}{3}\,I_8 - 7\,I_9 - \tfrac{8}{3}\,I_{11} 
- \tfrac{58}{3}\,I_{12} + \nabla_\alpha\,(-5\,J_{(1)}^\alpha 
+ 2\,J_{(2)}^\alpha)\; ,\\
I_5 &= \tfrac{1}{5}\,I_2 + \tfrac{14}{3}\,I_6 + \tfrac{4}{3}\,I_7 
+ \tfrac{1}{3}\,I_8 - 4\,I_9 - \tfrac{4}{3}\,I_{11} 
- \tfrac{32}{3}\,I_{12} - 4\,\nabla_\alpha\,J_{(1)}^\alpha\;.
\end{align}
As verified in Ref.~\cite{Boulanger:2025oli}, the action \eqref{8dConfActionEinstein} 
is equivalent to (i.e., it gives the same equations of motion as) 
the eight-dimensional $Q$-curvature, first computed in Ref.~\cite{Gover:2002ay}. 
Explicitly, one has~\cite{Boulanger:2025oli}
\begin{equation}\label{S8andQ8}
I_{\rm BR}[g_{\mu\nu}] = \tfrac{576 \pi^4}{5}\,\chi(M_8)
+ \tfrac{3}{20}\,\int_{M_8}\diff^8 x\,\,{\cal Q}_8 \;, 
\end{equation}
where total derivatives in ${\cal Q}_8\,$ are discarded.
In four dimensions, the analogous result between the 
topologically renormalised \cite{Anastasiou:2016jix}
Einstein-Hilbert action with negative cosmological constant 
and the Q-curvature is, dropping total derivatives in ${\cal Q}_4\,$,
\begin{equation}
-\tfrac{1}{4}\,\int_{M_4}\diff^4 x\,{\sqrt{-g}}\,W_{\mu\nu\rho\sigma}
W^{\mu\nu\rho\sigma} = 8\pi^2\,\chi(M_4) + \int_{M_4}\diff^4 x\,{\cal Q}_4\;.
\label{DecompositionQ4}
\end{equation}
In six dimensions, the L{\"u}-Pang-Pope conformal gravity 
action $I_{LPP}[g_{\mu\nu}]$ \cite{Lu:2013hx} that was shown in \cite{Anastasiou:2018mfk}
to renormalise Einstein-Hilbert's action with negative cosmological 
constant, is related to the Q-curvature 
as follows\footnote{In \eqref{Euler} we use a different phase 
convention for the Euler characteristic as compared to \cite{Boulanger:2025oli}.}
\begin{equation}
I_{LPP}[g_{\mu\nu}] = -128\pi^3\,\chi(M_6) \,- \,2\int_{M_6} 
\diff^6 x\;{\cal Q}_6\;.
\end{equation}
The variation of the $Q$-curvature gives the obstruction 
tensor \cite{graham2005ambient}, 
which is a symmetric, divergenceless and traceless rank-two tensor 
of conformal dimension $-6$ (in eight dimensions), 
that vanishes on conformally Einstein manifolds. 
The explicit expression of the eight-dimensional obstruction tensor 
was computed in Ref.~\cite{gover2006ambient}.

Now, we consider on the one hand the conformal gravity action with Einstein sector given in Eq.~\eqref{J8dConfActionEinstein}, up to an overall constant $\alpha$, supplemented by an arbitrary combination of the two independent divergences of Eqs.~\eqref{J1} and \eqref{J2} with parameters $\beta$ and $\gamma$, given by
\begin{equation}\label{L8prime}
\mathcal{L}'_8(\alpha,\beta,\gamma) := \alpha\,\mathcal{L}_8 + \beta\,\nabla_\mu\,J^\mu_{(1)} + \gamma\,\nabla_\mu\,J^\mu_{(2)}\,.
\end{equation}
On the other hand, we consider the Einstein-AdS Lagrangian in eight dimensions supplemented by both the topological Euler density (as in Eq.~\eqref{TopRen8d}) and the three possible boundary terms \eqref{BndTerms} with arbitrary free coefficients $c_1, c_2, c_3$, i.e.,
\begin{equation}\label{ActionHEEbnd}
\mathcal{L}_{\text{ren}}(c_1,c_2,c_3) := \tfrac{\ell^6}{1280\pi G}\,(\tfrac{1}{\ell^4}\,\Phi_{(2)}-\tfrac{1}{36\,\ell^2}\,\Phi_{(3)}+ \tfrac{1}{576}\,\Phi_{(4)}
+ c_1\,\Box^2\,\Phi_{(2)}
+ c_2\,\Box\,\Phi_{(3)} 
+ \tfrac{c_3}{\ell^2}\,\Box\,\Phi_{(2)})\,.
\end{equation}
We want to verify whether there is a choice of the six free parameters $\alpha, \beta, \gamma, c_1, c_2, c_3$ such that the two expressions given in Eqs.~\eqref{L8prime} and \eqref{ActionHEEbnd} are equal when evaluated on Einstein-AdS manifolds, i.e.,
\begin{equation}
\left.\mathcal{L}'_8(\alpha,\beta,\gamma)\right|_E = \left.\mathcal{L}_{\text{ren}}(c_1,c_2,c_3)\right|_E\,.
\end{equation}
We find that there is a unique choice of the parameters ensuring the equality, i.e.,
\begin{equation}\label{CoeffMatching}
\alpha = -\frac{\ell^6}{504 \pi G}\,, \quad
\beta = \frac{\ell^6}{360 \pi G}\,, \quad
\gamma = -\frac{7 \ell^6}{1080 \pi G}\,, \quad
c_1 = \frac{1}{72}\,, \quad
c_2 = -\frac{1}{216}\,, \quad
c_3 = \frac{1}{4}\,.
\end{equation}
Remarkably, the values $c_1, c_2, c_3$ are precisely those found 
in Eq.~\eqref{BulkBndAction}, which ensure that the whole expression 
\eqref{ActionHEEbnd} is finite.

Moreover, as explained in detail in Appendix \ref{VanishBndConf}, 
the asymptotic falloff of the total derivative conformal invariant densities is of order $z^9$ for Einstein-AdS manifolds.
Therefore, because they are total derivatives and the boundary is at $z\rightarrow 0$, 
their integral is identically zero for Einstein-AdS manifolds, i.e.
\begin{equation}
\int\limits_{\mathcal{M}_8}\left.d^8 x\,\sqrt{-g}\,\nabla_\mu\,J^\mu_{(i)}\right|_E=0\,,\quad  i=1,2\,.
\end{equation}
Therefore the conclusion is that, for Einstein-AdS manifolds, the renormalised Einstein-AdS action \eqref{BulkBndAction}
is exactly the action \eqref{8dConfActionEinstein}, without taking into account the total derivative conformal invariant densities obtained from Eqs.~\eqref{J1} and \eqref{J2}, i.e.
\begin{equation}
\left.I_{\rm ren}\left[g_{\mu \nu}\right]\right|_{E} = -\frac{\ell^6}{504 \pi G}\,\left.I_{\rm BR}[g_{\mu\nu}]\right|_E\,.
\end{equation}
This is in agreement with the fact that said conformal invariants cannot be fixed from the equation of motion condition of having an Einstein sector, 
and also with the fact that, as they are in themselves conformal invariants, they cannot be obtained from the conformal completion of bulk objects.

\section{Conclusion and outlook}

Building on the results in 4D, 6D \cite{Anastasiou:2016jix,Anastasiou:2020mik},
and now also in 8D, we conjecture that conformal invariance and the admissibility of an 
Einstein-AdS sector reproduce the holographically renormalised 
Einstein-Hilbert action with negative cosmological constant, 
in every even-dimensional, conformally compact manifold.

In an odd-dimensional, conformally compact Einstein 
manifold $(X_{d+1},g_+)$, the on-shell, regularised Einstein-Hilbert 
action with negative cosmological constant can be computed following 
Holographic Renormalisation  \cite{deHaro:2000vlm}, 
a method that is defined for both odd and even dimensional manifolds. At the same time, the on-shell value of the Einstein-Hilbert 
action with cosmological constant diverges like the volume of the 
manifold.
In an odd-dimensional bulk, it is known \cite{Graham:1999jg} 
that the renormalised volume is ambiguous. 
Indeed, the finite term in the asymptotic expansion of the volume 
depends on the representative metric in the conformal class at 
the conformal boundary. This ambiguity is understood from the 
AdS/CFT point of view \cite{Balasubramanian:1999re}.
In the paper \cite{Graham:1999jg}, Graham proved (Theorem 3.1.) 
that the renormalised volume of an even-dimensional, conformally 
compact manifold is independent of the choice of representative 
metric $g_{(0)}$ in the conformal class at the boundary. 
He further commented that the renormalised volume in that case 
is (we quote) an \emph{absolute invariant} of the conformally 
compact Einstein metric in the interior of the conformally compact 
manifold. That ``absolute'' is synonymous with ``global conformal" 
was made explicit somewhat later, in   
Refs.~\cite{Albin:2005qka,Chang:2005ska}.
The conformal invariance of the renormalised volume was
checked explicitly, from the holographic point of view, 
in 4D and 6D \cite{Anastasiou:2016jix,Anastasiou:2018mfk}, and now in 8D in 
the present paper as well.

A posteriori, the polynomial in the tensor $\Omega$ in
Eq.~\eqref{polynomialWE} represents the second-derivative 
sector of a higher-derivative gravity theory. 
Therefore, one might try to perform a conformal completion/covariantisation 
of the Einstein-AdS action. While this is direct in four dimensions 
\cite{Anastasiou:2016jix}, such construction in six dimensions requires 
the use of a suitable basis for the conformal invariants 
\cite{Anastasiou:2020mik,Anastasiou:2021tlv}. 
We would like to extend this procedure to the 8D case and, possibly, 
to generalise it to an arbitrary even dimension in the bulk.

In both four and six-dimensional cases, surface/energy functionals 
were derived from the Conformal Gravity action \cite{Anastasiou:2022ljq,Anastasiou:2024rxe}
vantage point. 
These structures appear as 
the singular contribution of a bulk gravity action evaluated on a manifold 
with a conical defect \cite{Lewkowycz:2013nqa}. They are conformally 
invariant themselves in codimension-2.
It would be interesting to relate these observations to the results 
presented in the Section 4 (see Theorem 4.1) of \cite{Graham:1999jg}, 
see also the companion paper \cite{Graham:1999pm}.
Within the framework of gauge/gravity duality, they correspond to 
a renormalised version of Holographic Entanglement Entropy and generalisations 
of it to non-minimal surfaces. In other words, they connect Quantum Information 
measures to conformal geometry, as notions such as Renormalized Area, 
Reduced Hawking Mass and Willmore energy can be readily reproduced 
\cite{Anastasiou:2022ljq,Anastasiou:2024rxe}. On the other hand, there is 
stacking evidence to support the idea that all of these codimension-2 
functionals would be finite for surfaces embedded in AAdS spacetimes 
\cite{Anastasiou:2025dex}. This idea extends the relation between
Renormalised bulk Volume and codimension-2 Area for minimal surfaces in AdS gravity to higher dimensions. 
The relation between the 8D Conformal Gravity with an Einstein 
sector and extremal surface functionals embedded therein will be sought for in future work.

\section*{Acknowledgements}

The work of N.B. was partially supported by the F.R.S.-FNRS PDR grant number T.0047.24. The work of D.R. is supported by Fondazione Angelo della Riccia grant 2026 and by DDFIP Essone. This work is partially funded by ANID FONDECYT (Chile) grants  11240059, 1240043, 1230492, 1231133, 1231779 and 1261016.  
G.A., I.J.A., R.O. and D.R. are grateful to the \emph{Physique de l'Univers, 
Champs et Gravitation} unit at UMONS for its hospitality during our visits (in particular during the workshop 
\href{https://web.umons.ac.be/pucg/en/event/workshop-conformal-higher-spins-twistors-and-boundary-calculus/}{\tt Conformal higher spins, twistors and 
boundary calculus}, UMONS, 30 June -- 4 July 2025), which allowed us to collaborate on the current project.

\appendix

\section{Definition of various useful tensors}
\label{Cotton et al}

In $d+1$ dimensions, the Schouten tensor is defined by
\begin{equation}
S_{\mu\nu} = \frac{1}{d-1}\,(R_{\mu\nu}-\frac{1}{2\,d}\,R\,g_{\mu\nu})\,.
\end{equation}
The Ricci tensor in terms of the Schouten one reads
\begin{equation}
R_{\mu\nu} = (d-1)\,S_{\mu\nu} + S\,g_{\mu\nu}\,,
\end{equation}
where $S$ is the trace of the Schouten tensor.
The Riemann tensor is decomposed in terms of its traceless part (Weyl tensor $W_{\mu\nu\rho\sigma}$) and of the Schouten tensor in the following way:
\begin{equation}
R_{\mu\nu\rho\sigma} = W_{\mu\nu\rho\sigma} + g_{\mu\rho}\,S_{\nu\sigma} - g_{\nu\rho}\,S_{\mu\sigma} - g_{\mu\sigma}\,S_{\nu\rho} + g_{\nu\sigma}\,S_{\mu\rho}\,.
\end{equation}
The Cotton tensor is defined by
\begin{equation}
C_{\mu\nu\rho} = \nabla_\rho\,S_{\mu\nu} - \nabla_\nu\,S_{\mu\rho}\,
\end{equation}
which can be written in terms of the divergence of the Weyl tensor, by means of the differential Bianchi identity $\nabla_{[\mu}\,R_{\nu\rho]\alpha\beta} = 0$:
\begin{equation}\label{Cotton}
C_{\mu\nu\rho} = -\frac{1}{d-2}\,\nabla^\sigma\,W_{\sigma\mu\nu\rho}\,.
\end{equation}
The Bach tensor is defined by
\begin{equation}
B_{\mu\nu} = S^{\rho\sigma}\,W_{\mu\rho\nu\sigma} + \nabla^\rho\,C_{\nu\mu\rho}\,.
\end{equation}

Einstein equations in $d+1$ dimensions with negative cosmological constant impose the Ricci tensor to be equal to 
\begin{equation}
R_{\mu\nu}\vert_E = -\frac{d}{\ell^2}\,g^{(E)}_{\mu\nu}\,.
\end{equation}
In the manifolds satisfying this condition (\emph{Einstein-AdS manifolds}), the
Schouten, Riemann, Cotton, and Bach tensors become respectively
\begin{equation}
S_{\mu\nu}\vert_E = -\frac{1}{2\,\ell^2}\,g^{(E)}_{\mu\nu}\,, \;\;
R_{\mu\nu}{}^{\rho\sigma}\vert_E = W_{\mu\nu}{}^{\rho\sigma} 
- \frac{1}{\ell^2}\,(\delta_{\mu}^{\rho}\delta_{\nu}^{\sigma} 
- \delta_{\mu}^{\sigma}\delta_{\nu}^{\rho})\,, \;\;
C_{\mu\nu\rho}\vert_E = B_{\mu\nu}\vert_E = 0\,.
\end{equation}

\section{Weyl-covariant tensor calculus}
\label{sec:Weyl-covariant}

In this appendix the Weyl-covariant tensor calculus developed in 
Ref.~\cite{Boulanger:2004eh} is briefly reviewed. 
We use the conventions and notation of Refs.~\cite{Boulanger:2007ab,Boulanger:2007st},
where the classification of Weyl anomalies in arbitrary dimension 
was obtained.
Under infinitesimal Weyl rescalings of the metric
\begin{equation}
    \delta_\sigma g_{\mu\nu} = 2\,\sigma(x)\,g_{\mu\nu}\;,
\end{equation}
the components of Weyl tensors are invariant:  
$\delta_\sigma W^\mu{}_{\nu\alpha\beta} = 0\,$.
Denoting by $\Delta_\mu{}^\nu\,$ the $GL(D)$ generators that act on tensors 
through $\Delta_\mu{}^\nu \,T^\alpha_\beta = \delta^\nu_\beta T^\alpha_\mu - 
\delta^\alpha_\mu T^\nu_\beta\,$, 
the symbol $\nabla_\mu = \partial_\mu - \Gamma_{\mu\nu}{}^\rho\,\Delta_\rho{}^\nu$ 
denotes the usual torsion-free metric-compatible 
(Levi-Civita) covariant derivative associated with the Christoﬀel symbols
$\Gamma_{\mu\nu}{}^\rho$, in terms of which 
$R^\mu{}_{\nu\rho\sigma} = \partial_{\rho}\Gamma_{\nu\sigma}{}^{\mu}+\ldots$.
The commutator of covariant derivatives gives  $[\nabla_\mu,\nabla_\nu]V^\rho=R^\rho{}_{\sigma\mu\nu}V^\sigma\,$ 
and, in general, 
$[\nabla_\mu,\nabla_\nu] = R_{\mu\nu\rho}{}^\sigma{}\,\Delta_\sigma{}^\rho\,$.
The components of the Cotton tensor are given by 
$C_{\alpha\rho\sigma} = 2\, \nabla_{[\sigma} S_{\rho]\alpha} \equiv 
\nabla_{\sigma} S_{\rho\alpha} - \nabla_{\rho} S_{\sigma\alpha}\,$.
The Weyl-covariant derivative constructed in Ref.~\cite{Boulanger:2004eh} 
is given by 
\begin{equation}
{\cal D}_{\mu} = \nabla_{\mu}+S_{\mu\alpha}\,\boldsymbol{\Gamma}^\alpha\;,
\end{equation}
where we refer to this work for the definition of 
the generators $\boldsymbol{\Gamma}^\alpha$; see also below for 
a few examples. Importantly, because 
$\boldsymbol{\Gamma}^\alpha(g_{\mu\nu})\equiv 0\,$,  
the metric is preserved by the Weyl-covariant derivative: 
${\cal D}_{\alpha}g_{\mu\nu}=0\,$.
The important property of the Weyl-covariant 
derivative $\cal D$ is that its curvature vanishes if and only if the 
metric is conformally flat. Explicitly, one has \cite{Boulanger:2004eh}
\begin{equation}
    [{\cal D}_\mu,{\cal D}_\nu] = 
      W_{\mu\nu\rho}{}^\sigma{}\,\Delta_\sigma{}^\rho
    - C_{\alpha\mu\nu}\,\boldsymbol{\Gamma}^\alpha\;.
\end{equation}
The first term on the right-hand side is the same as in the expression 
for the commutator of the Levi-Civita covariant derivative, 
except that now the Weyl tensor replaces the Riemann curvature tensor.
The second term on the right-hand side brings the Cotton tensor, which is 
the conformal field strength in 3D, where the Weyl tensor identically vanishes.
In dimensions $d+1>3\,$, the Cotton tensor can be written as a covariant divergence 
of the Weyl tensor, viz., 
$C_{\alpha\rho\sigma} = -\frac{1}{d-2}\,\nabla_\mu W^\mu{}_{\alpha\rho\sigma}\,$. 

Similarly to the fact that the tensors in (pseudo)Riemann 
geometry are given by the metric tensor, the Riemann tensor, all its 
covariant derivatives and traces thereof using the (inverse)metric tensor, 
the set of \emph{$W$-tensors} is given by the Weyl tensor, all its Weyl-covariant 
derivatives and their non-trivially vanishing traces. 
We introduce super indices and the notation
\[\{W_{\Omega_0}, W_{\Omega_1}, \ldots, W_{\Omega_k}, \ldots\}
=\{W^\mu{}_{\nu\rho\sigma}, {\cal D}_{\alpha_1} W^\mu{}_{\nu\rho\sigma},
\ldots, {\cal D}_{\alpha_k} {\cal D}_{\alpha_{k-1}}\ldots {\cal D}_{\alpha_1} W^\mu{}_{\nu\rho\sigma},\ldots\}\;.\]
The defining property of the $W$-tensors is that they transform, under infinitesimal 
Weyl rescalings of the metric, 
with the first derivative of the Weyl parameter only \cite{Boulanger:2004eh}:
\begin{equation}
 \delta_\sigma W_{\Omega_i} = \partial_\alpha\sigma \,
 [{T}^{\alpha}]_{\Omega_i}{}^{\Omega_{i-1}}\,W_{\Omega_{i-1}}\;.   
\end{equation}
After introducing the tensor 
${\cal P}^{\alpha\nu}_{\mu\beta}:=-g^{\alpha\nu}g_{\mu\beta}
+\delta^\alpha_\mu\delta^\nu_\beta+\delta^\alpha_\beta\delta^\nu_\mu\,$, 
we can present the first few $W$-tensors as follows: 
\begin{align}
W_{\Omega_0} :=~& W^\mu{}_{\nu\rho\sigma}\;,
\nonumber \\
W_{\Omega_1} :=~& {\cal D}_{\alpha_1}\,W^\mu{}_{\nu\rho\sigma} 
\equiv {\nabla}_{\alpha_1}\,W^\mu{}_{\nu\rho\sigma}\;,
\nonumber \\
W_{\Omega_2}  :=~&  
{\cal D}_{\alpha_2}{\cal D}_{\alpha_1} W^\mu{}_{\nu\rho\sigma}
\equiv \nabla_{\alpha_2}\,\nabla_{\alpha_1}\,W^\mu{}_{\nu\rho\sigma} 
- S_{\alpha_2 \lambda}\,{\cal P}^{\lambda\delta}_{\epsilon\alpha_1}\,
\Delta_\delta{}^\epsilon\,W^\mu{}_{\nu\rho\sigma}\;,
\nonumber \\
W_{\Omega_3} :=~& 
{\cal D}_{\alpha_3}{\cal D}_{\alpha_2}{\cal D}_{\alpha_1} 
W^\mu{}_{\nu\rho\sigma}
\equiv \nabla_{\alpha_3}\,
{\cal D}_{\alpha_2}{\cal D}_{\alpha_1}W^\mu{}_{\nu\rho\sigma} 
\nonumber \\
& -S_{\lambda\alpha_3}\,(\delta^\gamma_{\alpha_1}\,
{\cal P}^{\lambda\delta}_{\epsilon\alpha_2}\,
\Delta_\delta{}^\epsilon + \delta^\gamma_{\alpha_2}\,
{\cal P}^{\lambda\delta}_{\epsilon\alpha_1}\,\Delta_\delta{}^\epsilon - 
{\cal P}^{\lambda\gamma}_{\alpha_1\alpha_2})\,
{\cal D}_{\gamma}W^\mu{}_{\nu\rho\sigma}\;,
\nonumber\\
W_{\Omega_4} :=~&
{\cal D}_{\alpha_4}{\cal D}_{\alpha_3}{\cal D}_{\alpha_2}{\cal D}_{\alpha_1} 
W^\mu{}_{\nu\rho\sigma}
\nonumber \\
&\equiv
\nabla_{\alpha_4}\,W^\mu{}_{\nu\rho\sigma,\alpha_1\alpha_2\alpha_3} 
- S_{\lambda\alpha_4}\,
\Big(\delta^{\gamma_1}_{\alpha_1}\,\delta^{\gamma_2}_{\alpha_2}\,
{\cal P}^{\lambda\delta}_{\epsilon\alpha_3}\,\Delta_\delta{}^\epsilon 
+ \delta^{\gamma_1}_{\alpha_1}\,\delta^{\gamma_3}_{\alpha_2}\,
{\cal P}^{\lambda\delta}_{\epsilon\alpha_2}\,\Delta_\delta{}^\epsilon 
\nonumber \\
& \quad + \delta^{\gamma_1}_{\alpha_2}\,\delta^{\gamma_2}_{\alpha_3}\,
{\cal P}^{\lambda\delta}_{\epsilon\alpha_1}\,\Delta_\delta{}^\epsilon 
- \delta^{\gamma_1}_{\alpha_2}\,{\cal P}^{\lambda\gamma_2}_{\alpha_1\alpha_3} 
- \delta^{\gamma_1}_{\alpha_1}\,{\cal P}^{\lambda\gamma_2}_{\alpha_2\alpha_3}- 
\delta^{\gamma_1}_{\alpha_3}\,
{\cal P}^{\lambda\gamma_2}_{\alpha_1\alpha_2}\Big)\,
{\cal D}_{\gamma_2}{\cal D}_{\gamma_1}W^\mu{}_{\nu\rho\sigma}\;.
\nonumber
\end{align}

\section{Vanishing of boundary conformal invariants in Einstein manifolds}
\label{VanishBndConf}

Due to the vanishing of the Cotton for Einstein spacetimes, the only surviving terms from \(J^\alpha_{(1)}\) and \(J^\alpha_{(2)}\) of Eqs.~\eqref{J1} and \eqref{J2} are
\begin{align}
J^\alpha_{(1)}
&=
W^{\alpha\beta\gamma\delta}
W_{\beta}{}^{\epsilon\sigma\rho}
\nabla_{\rho}W_{\gamma\delta\epsilon\sigma},
\label{eq:J1-surviving}
\\[0.75em]
J^\alpha_{(2)}
&=
W^{\alpha\beta\gamma\delta}
W_{\gamma}{}^{\epsilon\sigma\rho}
\nabla_{\rho}W_{\beta\epsilon\delta\sigma}
-
W^{\alpha\beta\gamma\delta}
W_{\beta}{}^{\epsilon\sigma\rho}
\nabla_{\rho}W_{\gamma\delta\epsilon\sigma}.
\label{eq:J2-surviving}
\end{align}
Also, as the conformal boundary is located at $z\rightarrow0$, one has that 
\begin{equation}\label{C3}
\int\limits_{\mathcal{M}_8}d^{8}x\sqrt{-g}\,\nabla_{\mu}J^{\mu}_{(i)}=\int\limits_{\mathcal{M}_8}d^{8}x\partial_{\mu}\left(\sqrt{-g}\,J^{\mu}_{(i)}\right)=\lim_{z\rightarrow0}\int\limits_{\partial\mathcal{M}_8}d^{7}x\sqrt{-g_{(0)}}\frac{\ell^8}{z^8}\left(1+\mathcal{O}\left(z^2\right)\right)J^{z}_{(i)}.
\end{equation}
Therefore, it is enough to analyse the asymptotic order in $z$ of $J^{z}_{(1)}$ and, mutatis mutandis, $J^{z}_{(2)}$.

Near the conformal boundary, considering the Fefferman-Graham metric given in Eq.~\eqref{gaussnormal}, one has that lowering an index reduces the z order by two and raising an index increases the z order by two. Also, for the radial lapse $N$ and extrinsic curvature $K$ one has that
\begin{equation}
 N=\frac{\ell}{z},
 \qquad
 K^i{}_j\sim \mathcal{O}(1).
\end{equation}
The relevant non-vanishing Christoffel symbols scale as
\begin{align}
& \Gamma^{i}{}_{jk}(h) = \Gamma^{i}{}_{jk}(\bar g) \sim \mathcal{O}(1)\,, \label{eq:Gamma-ijk}\\
&\Gamma^{z}{}_{ij} = \frac{1}{N}\,K_{ij} \sim \mathcal{O}(z^{-1})\,,\label{eq:Gamma-zij}\\
&\Gamma^{i}{}_{jz} = - N\,K^i{}_j \sim \mathcal{O}(z^{-1})\,,\label{eq:Gamma-ijz}\\
&\Gamma^{z}{}_{zz} = \frac{N'}{N} = -\frac{1}{z} \,.\label{eq:Gamma-zzz}
\end{align}
The Weyl components needed for the power-counting are given in Eq.~\eqref{WeylAsymptotic}. Finally, consider that after raising or lowering indices with the bulk metric, it turns out that each extra radial index in a Weyl tensor increases the power of
\(z\) by one relative to the corresponding all-boundary component.
Starting from
\begin{align}
J^z_{(1)}=
W^{z\beta\gamma\delta}\,W_{\beta}{}^{\epsilon\sigma \rho}
\Big(&\,\partial_ {\rho} W_{\gamma\delta\epsilon\sigma}
-\Gamma^{\lambda}{}_{\rho \gamma}W_{\lambda\delta\epsilon\sigma}
-\Gamma^{\lambda}{}_{\rho \delta}W_{\gamma\lambda\epsilon\sigma}
\nonumber\\
&-\Gamma^{\lambda}{}_{\rho \epsilon}W_{\gamma\delta\lambda\sigma}
-\Gamma^{\lambda}{}_{\rho \sigma}W_{\gamma\delta\epsilon\lambda}\Big), 
\end{align}
we identify the contributions $\rho=z$ and $\rho = i$.
In the first case, the leading order fall-off of the radial covariant derivative is
\begin{equation}
\nabla_z W_{\gamma\delta\epsilon\sigma}
\sim \mathcal{O}(z^{-3})\,.
\end{equation}
Therefore,
\begin{equation}
W^{z\beta\gamma\delta}\,W_{\beta}{}^{\epsilon\sigma z}\,\nabla_z W_{\gamma\delta\epsilon\sigma}
\sim \mathcal{O}(z^7)\,\mathcal{O}(z^5)\,\mathcal{O}(z^{-3}) =
\mathcal{O}(z^9)\,.\label{eq:rho-z-final}
\end{equation}
The estimate
\(
\nabla_z W_{\gamma\delta\epsilon\sigma}\sim \mathcal{O}(z^{-3})
\)
is saturated already for the all-boundary component
\(
\nabla_zW_{ijkl}\sim \mathcal{O}(z^{-3})
\), because
\(
\partial_z W_{ijkl}
\)
and the connection terms containing
\(
\Gamma^a{}_{zi}W_{ajkl}
\)
can both behave as
\(
\mathcal{O}(z^{-3})
\).

Now, for the second case of $\rho=i$, there are three leading contributions to $J^z_{(1)}$ corresponding to $W^{zjkl}W_{j}{}^{mn i}\nabla_i W_{klmn}$, $W^{zjkl}W_{j}{}^{zm i}\nabla_i W_{zklm}$ and $W^{zj zk}W_{j}{}^{lm i}\nabla_i W_{zklm}$. Then, because $\nabla_i W_{klmn}
\sim \mathcal{O}(z^{-2})$ and $\nabla_i W_{zklm}
\sim \mathcal{O}(z^{-3})$, one sees that all the contributions are of order $\mathcal{O}(z^{9})$. Consequently, at leading order this term falls-off as
\begin{equation}
W^{z\beta\gamma\delta}\,W_{\beta}{}^{\epsilon\sigma i}\,\nabla_i W_{\gamma\delta\epsilon\sigma}
\sim
\mathcal{O}(z^9).
\label{eq:rho-i-final}
\end{equation}
Note that the covariant derivative with all radial indices 
$\nabla_i W_{jkln}$ also includes terms such as 
$\Gamma^z_{ij} W_{zkln}$, which are still overall $\mathcal{O}\left(z^{-2}\right)$, although $\Gamma^z_{ij}\sim\mathcal{O}\left(z^{-1}\right)$, due to the improved asymptotic behaviour of $W_{zkln}\sim\mathcal{O}\left(z^{-1}\right)$.

Finally, since $J^z_{(2)}$ contains the same type of contractions of the same objects as $J^z_{(1)}$, one concludes that it has the same falloff. Overall, one obtains that
\begin{equation}
J^z_{(i)}
\sim \mathcal{O}(z^9)\,,
\end{equation}
and through Eq.~\eqref{C3}, it can be concluded that for Einstein-AdS manifolds the integral of the conformal boundary densities vanishes, i.e.,
\begin{equation}
\int\limits_{\mathcal{M}_8}\left.\text{d}^8 x\,\sqrt{-g}\,\nabla_\mu\,J^\mu_{(i)}\right|_E=0\,,\quad  i=1,2\,.
\end{equation}

\bibliographystyle{JHEP}
\bibliography{biblio-2}

\end{document}